\documentclass[preprint,showpacs,preprintnumbers,amsmath,amssymb]{revtex4-1}
\usepackage{graphicx}
\usepackage{dcolumn}
\usepackage{bm}
\usepackage{amssymb}

\newcommand{\mathe}{\mathrm{e}}

\begin{document}

\title{Synchronization of Phase-coupled Oscillators with Distance-dependent Delay}
\author{Karol Trojanowski}
\email[e-mail address:]{karol.trojanowski@uj.edu.pl}
\affiliation{Marian Smoluchowski Institute of Physics, Department
of Statistical Physics and Mark Kace Center for Complex Systems Research, Jagiellonian University, Reymonta 4,
Krak\'ow, Poland}
\author{Lech Longa}
\email[e-mail address:]{lech.longa@uj.edu.pl}
\affiliation{Marian Smoluchowski Institute of Physics, Department
of Statistical Physics and Mark Kace Center for Complex Systems Research, Jagiellonian University, Reymonta 4,
Krak\'ow, Poland}

\date{\today}

\begin{abstract}
  By means of numerical integration we investigate the coherent and incoherent
  phases in a~generalized Kuramoto model of phase-coupled oscillators
  with distance-depen{\nobreak}dent delay. Preserving the topology of a
  complete graph, we arrange the nodes on a square lattice while introducing
  finite interaction velocity, which gives rise to non-uniform delay. It is
  found that such delay facilitates incoherence and removes reentrant behavior
  found in models with uniform delay. A~coupling-delay phase diagram is
  obtained and compared with previous results for uniform delay.
\end{abstract}

\maketitle

\section{Model summary}

\subsection{The Kuramoto model}

The popular Kuramoto model of mutual synchronization of coupled oscillators
{\cite{kuramoto-cite}} has, since its inception, drastically improved the
understanding of this prevalent phenomenon. Common examples
{\cite{acebron}}{\cite{from-kuramoto-to-crawford}} include synchronous
chirping of crickets, flashing of Chinese fireflies {\cite{fireflies}},
clapping of audiences, bursting of neurons, contraction of heart muscles or
operation of Josephson junction arrays
{\cite{josephson-1980}}{\cite{josephson-strogatz}} to name a few. This model still remains
the most succesful one, due to its mathematical tractability, combined with
the ability to capture the essence of synchrony.

We build up from the definition of the Kuramoto model which is most suitable
for direct treatment by numerical methods {\cite{acebron}}{\cite{daniels}}:
\begin{equation}
  \dot{\theta}_i (t) = \omega_i + \frac{K}{N} \sum_{j = 1}^N \sin (\theta_j
  (t) - \theta_i (t)) \label{kuramoto}
\end{equation}
where $i = 1 \ldots N$, $\theta_i (t)$ is the phase of the $i$-th oscillator
at time $t$ and $\omega_i$ are intrinsic oscillator frequencies, sampled from yet unspecified probability distribution $\rho (\omega)$ on compact
support. Kuramoto solved this model exactly in the case of $N \rightarrow
\infty$ and $\omega_i$ sampled from a Lorentz distribution. Solutions for
other distributions have subsequently been obtained. A~model such defined
exhibits a (mean-field-type) phase transition between the disordered
(incoherent) and ordered (coherent) phases as the coupling constant $K$ is
increased. Order is monitored by the real parameter $r$, defined as:
\begin{equation}
  r (t) \mathe^{i \psi (t)} = \frac{1}{N} \sum_{j = 1}^N \mathe^{i \theta_i
  (t)} . \label{r}
\end{equation}
When the stationary state is assumed, $r (t) = r$ ($r \epsilon [0, 1]$),
with $r = 1$ and $r = 0$ in total coherence and incoherence, respectively.

\subsection{Introducing delay}

Some real systems cannot be considered without taking delay into account. The
popular example of a clapping audience synchronizing to clap in unison is
valid only for sufficiently small audiences, such as opera halls. When
distances are of the order of $300 m$, or higher, the finite speed of sound makes
the delay non-negligible. As a result, e.g. football arena audiences cannot
clap together or have difficulty in coherent singing.

We start by introducing delay to (\ref{kuramoto}) in the most general way:
\begin{equation}
  \dot{\theta}_i (t) = \omega_i + \frac{K}{N} \sum_{j = 1}^N \sin (\theta_j (t
  - \tau_{ij}) - \theta_i (t)) . \label{kuramoto delayed}
\end{equation}
The case of uniform delay, $\tau_{ij} \equiv \tau$, is interpreted as coupling
of the state at $t$ to the state at $t - {\nobreak} \tau$. The stability of
incoherence in such a model has been studied by Yeung and Strogatz in
{\cite{strogatz-delay}}.

To introduce non-uniform delay, we arrange the nodes on a square lattice
while preserving the topology of a complete graph. The coupling remains
uniform, however the delay is made distance-dependent through the definition:
\begin{equation}
  \tau_{ij} = \tau \cdot \frac{s_{ij}}{\left\langle s_{ij} \right\rangle} .
  \label{delay}
\end{equation}
$\tau$ is interpreted as the inverse velocity. The distance $s_{ij}$ is
defined with the so-called ``taxi-driver's measure'', i.e. as sum of the
differences in horizontal and vertical coordinates and is measured in number
of nodes. To maintain translational invariance we identify the opposite edges
and the shortest route is always preferred. Hence, when $N = L \times L$, the
average distance between any pair of nodes $\left\langle s_{ij} \right\rangle
= \frac{L}{2}$. This definition normalizes the maximum delay to $2 \tau$ and
removes dependence on network size. We find that the unmodified parameter $r$,
defined by (\ref{r}), is useful in monitoring the average order in the sample.

\section{Simulations and results}

We have investigated the behavior of the model described by (\ref{kuramoto
delayed}) and (\ref{delay}) by integrating the equations (\ref{kuramoto
delayed}) using a four-step Adams-Bashforth scheme. For simplicity and
reference with previous results {\cite{strogatz-delay}}, we set $\omega_i =
\frac{\pi}{2}$ for all $i$, therefore $\rho (\omega) = \delta (\omega -
\frac{\pi}{2})$. A run for one pair of parameters $(\tau, {\nobreak} K)$
consisted of 10000 integration steps with step size $\Delta t = 0.01$, out of
which the last 6000 were considered for averaging the order parameter $r$ to
obtain the temporal average $\overline{r}$, rejecting the first 4000 when the
system is approaching stability. Lattices as large as $32 \times 32$ were
considered. The initial conditions, as well as histories of $\theta_i$, were
sampled uniformly from $[0, 2 \pi)$.

\begin{figure}
  a)\includegraphics[scale=0.4]{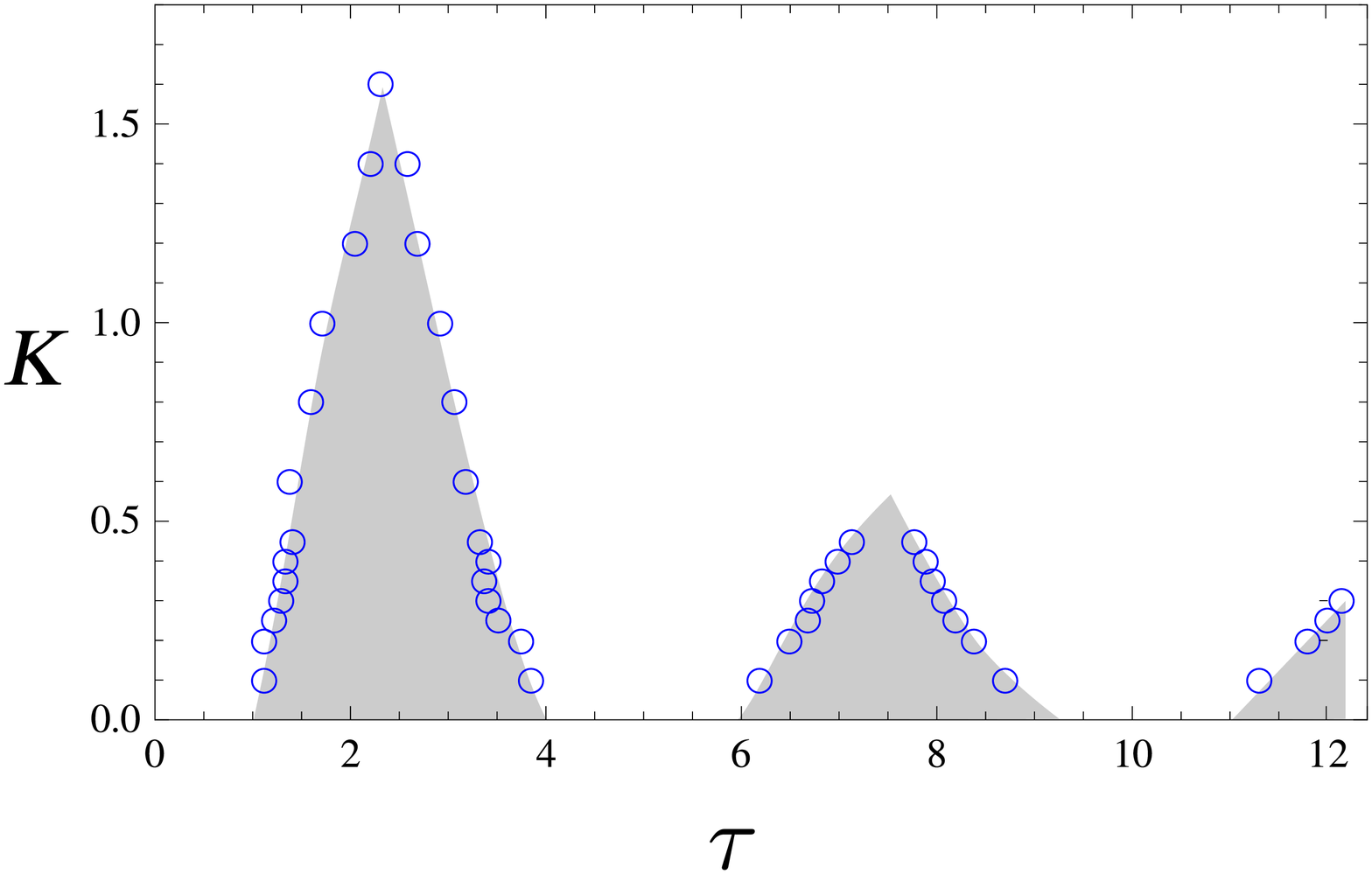}
  
  b)\includegraphics[scale=0.4]{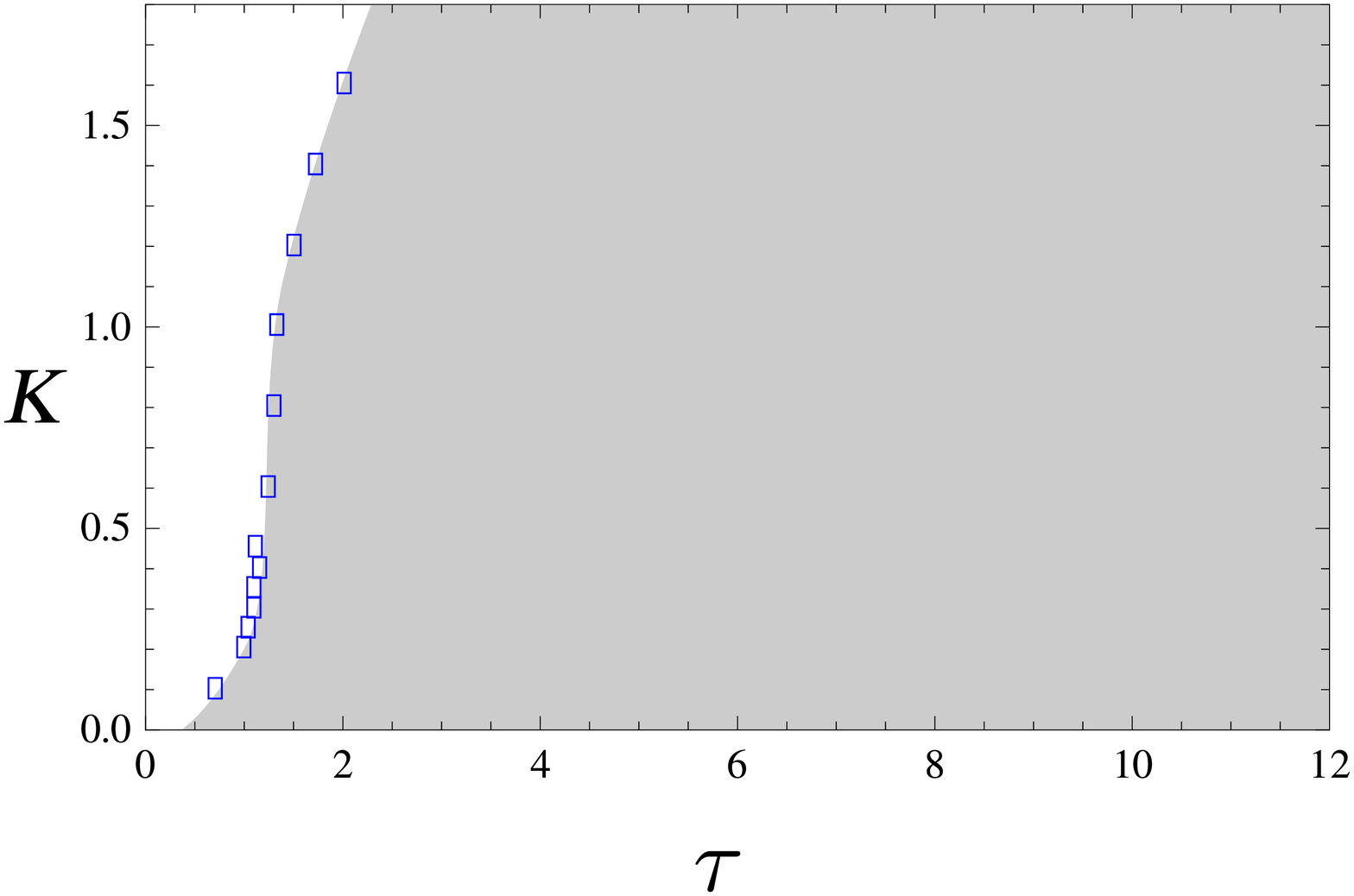}

  \caption{
	\label{unifig}Phase diagram portions for the uniform delay model
  (a) and the model described by (\ref{delay}) (b). The shaded areas visually
  approximate the incoherent regimes. The phase border points are results of
  numerical integration of (\ref{kuramoto delayed}).}
\end{figure}

We have found phase boundaries between the completely ordered and disordered
phase (Fig.~\ref{unifig}b). For reference, we have produced a diagram for
$\tau_{ij} \equiv \tau$ (Fig.~\ref{unifig}a). In the case of $\rho (\omega) =
\delta (\omega - \omega_0)$, the transitions occuring with changing $\tau$ are
instanteous. It is observed that non-uniform delay removes the reentrance of
synchrony, as intuitively expected. This difference is easily understood when
considering low coupling. The reentrance in the case of uniform delay and $K
\ll 1$ is due to there being sufficiently little difference between states at
$t - \tau$ and $t$ for the effect of delay to be approximated by the rotation
of all oscillators with average frequency:
\begin{equation}
  \sin (\theta_j (t - \tau) - \theta_i (t)) \approx \sin (\theta_j (t) -
  \theta_i (t) - \overline{\omega} \tau)
\end{equation}
The reentrance for low coupling then occurs when $\overline{\omega} \tau$ is
close to an integer multiple of $2 \pi$ ($\tau \approx k \frac{2
\pi}{\overline{\omega}}$, $k = 1, 2, \ldots$), where the low-delay limit is
reproduced. Our simulations reflect this heuristic quantitatively up to $k =
2$ and qualitatively from $k = 3$ on. In the case of distance-dependent delay
(Fig.~\ref{unifig}b) and $K \ll 1$ the effects of delay are individually
approximated by phase shifts of $\overline{\omega} \tau
\frac{s_{ij}}{\left\langle s_{ij} \right\rangle}$ which vary across
connections and the low-delay limit cannot be reproduced by a specific value
of delay. Hence, no reentrance occurs.

\section{Conclusions}

Our results prove that in order to realistically reproduce behavior of
synchronizable systems in which delay cannot be neglected, the
dependence of delay on distance must also be accounted for. However, only this
dependence was considered in this research, namely the case that even distant
nodes get to affect other distant nodes with the same strength as they affect
their closest neighbors, in consistence with the mean-field approximation. 
This is indeed true for some systems, such as digital
communication networks. The effect on the phase diagram when coupling
decreasing with distance is taken into account needs to be considered in due
course.

\begin{acknowledgements}
Project operated within the Foundation for Polish Science International PhD
Projects Programme co-financed by the European Regional Development Fund
covering, under the agreement no. MPD/2009/6, the Jagiellonian University
International PhD Studies in Physics of Complex Systems.
\end{acknowledgements}

\bibliographystyle{apsrev4-1}

\begin{thebibliography}{1}%
\makeatletter
\providecommand \@ifxundefined [1]{%
 \ifx #1\undefined \expandafter \@firstoftwo
 \else \expandafter \@secondoftwo
\fi
}%
\providecommand \@ifnum [1]{%
 \ifnum #1\expandafter \@firstoftwo
 \else \expandafter \@secondoftwo
\fi
}%
\providecommand \enquote [1]{``#1''}%
\providecommand \bibnamefont  [1]{#1}%
\providecommand \bibfnamefont [1]{#1}%
\providecommand \citenamefont [1]{#1}%
\providecommand\href[0]{\@sanitize\@href}%
\providecommand\@href[1]{\endgroup\@@startlink{#1}\endgroup\@@href}%
\providecommand\@@href[1]{#1\@@endlink}%
\providecommand \@sanitize [0]{\begingroup\catcode`\&12\catcode`\#12\relax}%
\@ifxundefined \pdfoutput {\@firstoftwo}{%
 \@ifnum{\z@=\pdfoutput}{\@firstoftwo}{\@secondoftwo}%
}{%
 \providecommand\@@startlink[1]{\leavevmode}%
 \providecommand\@@endlink[0]{}%
}{%
 \providecommand\@@startlink[1]{%
  \leavevmode
  \pdfstartlink
   attr{/Border[0 0 1 ]/H/I/C[0 1 1]}%
   user{/Subtype/Link/A<</Type/Action/S/URI/URI(#1)>>}%
  \relax
 }%
 \providecommand\@@endlink[0]{\pdfendlink}%
}%
\providecommand \url  [0]{\begingroup\@sanitize \@url }%
\providecommand \@url [1]{\endgroup\@href {#1}{\urlprefix}}%
\providecommand \urlprefix [0]{URL }%
\providecommand \Eprint[0]{\href }%
\@ifxundefined \urlstyle {%
  \providecommand \doi [1]{doi:\discretionary{}{}{}#1}%
}{%
  \providecommand \doi [0]{doi:\discretionary{}{}{}\begingroup
  \urlstyle{rm}\Url }%
}%
\providecommand \doibase [0]{http://dx.doi.org/}%
\providecommand \Doi[1]{\href{\doibase#1}}%
\providecommand \bibAnnote [3]{%
  \BibitemShut{#1}%
  \begin{quotation}\noindent
    \textsc{Key:}\ #2\\\textsc{Annotation:}\ #3%
  \end{quotation}%
}%
\providecommand \bibAnnoteFile [2]{%
  \IfFileExists{#2}{\bibAnnote {#1} {#2} {\input{#2}}}{}%
}%
\providecommand \typeout [0]{\immediate \write \m@ne }%
\providecommand \selectlanguage [0]{\@gobble}%
\providecommand \bibinfo [0]{\@secondoftwo}%
\providecommand \bibfield [0]{\@secondoftwo}%
\providecommand \translation [1]{[#1]}%
\providecommand \BibitemOpen[0]{}%
\providecommand \bibitemStop [0]{}%
\providecommand \bibitemNoStop [0]{.\EOS\space}%
\providecommand \EOS [0]{\spacefactor3000\relax}%
\providecommand \BibitemShut [1]{\csname bibitem#1\endcsname}%
\bibitem{kuramoto-cite}%
  \BibitemOpen
  \bibfield{author}{%
  \bibinfo {author} {\bibfnamefont{Y.}~\bibnamefont{{Kuramoto}}},\ }%
  \emph{\bibinfo {title} {Chemical Oscillations, Waves and Turbulence}}\
  (\bibinfo {publisher} {Springer Verlag},\ \bibinfo {address} {New York},\
  \bibinfo {year} {1984})%
  \bibAnnoteFile{NoStop}{kuramoto-cite}%
\bibitem{acebron}%
  \BibitemOpen
  \bibfield{author}{%
  \bibinfo {author} {\bibfnamefont{J.~A.}\ \bibnamefont{{Acebron}}}, \bibinfo
  {author} {\bibfnamefont{L.}~\bibnamefont{{Bonilla}}}, \bibinfo {author}
  {\bibfnamefont{C.~J.}\ \bibnamefont{{Perez Vincente}}}, \bibinfo {author}
  {\bibfnamefont{F.}~\bibnamefont{{Ritort}}},\ and\ \bibinfo {author}
  {\bibfnamefont{R.}~\bibnamefont{{Spigler}}},\ }%
  \bibfield{journal}{%
  \bibinfo {journal} {Review of Modern Physics}\ }%
  \textbf{\bibinfo {volume} {77}},\ \bibinfo {pages} {137} (\bibinfo {month}
  {January}\ \bibinfo {year} {2005})%
  \bibAnnoteFile{NoStop}{acebron}%
\bibitem{from-kuramoto-to-crawford}%
  \BibitemOpen
  \bibfield{author}{%
  \bibinfo {author} {\bibfnamefont{S.~H.}\ \bibnamefont{Strogatz}},\ }%
  \bibfield{journal}{%
  \bibinfo {journal} {Physica D}\ }%
  \textbf{\bibinfo {volume} {143}},\ \bibinfo {pages} {1} (\bibinfo {year}
  {2000})%
  \bibAnnoteFile{NoStop}{from-kuramoto-to-crawford}%
\bibitem{fireflies}%
  \BibitemOpen
  \bibfield{author}{%
  \bibinfo {author} {\bibfnamefont{J.}~\bibnamefont{Buck}}\ and\ \bibinfo
  {author} {\bibfnamefont{E.}~\bibnamefont{Buck}},\ }%
  \bibfield{journal}{%
  \bibinfo {journal} {Scientific American}\ }%
  \textbf{\bibinfo {volume} {234}},\ \bibinfo {pages} {74} (\bibinfo {month}
  {May}\ \bibinfo {year} {1976})%
  \bibAnnoteFile{NoStop}{fireflies}%
\bibitem{josephson-1980}%
  \BibitemOpen
  \bibfield{author}{%
  \bibinfo {author} {\bibfnamefont{N.~F.}\ \bibnamefont{{Pedersen}}}, \bibinfo
  {author} {\bibfnamefont{O.~H.}\ \bibnamefont{{Soerensen}}}, \bibinfo {author}
  {\bibfnamefont{B.}~\bibnamefont{{Dueholm}}},\ and\ \bibinfo {author}
  {\bibfnamefont{J.}~\bibnamefont{{Mygind}}},\ }%
  \bibfield{journal}{%
  \bibinfo {journal} {Journal of Low Temperature Physics}\ }%
  \textbf{\bibinfo {volume} {38}},\ \bibinfo {pages} {1} (\bibinfo {month}
  {January}\ \bibinfo {year} {1980})%
  \bibAnnoteFile{NoStop}{josephson-1980}%
\bibitem{josephson-strogatz}%
  \BibitemOpen
  \bibfield{author}{%
  \bibinfo {author} {\bibfnamefont{K.~A.}\ \bibnamefont{{Wiesenfeld}}},
  \bibinfo {author} {\bibfnamefont{P.}~\bibnamefont{{Colet}}},\ and\ \bibinfo
  {author} {\bibfnamefont{S.~H.}\ \bibnamefont{{Strogatz}}},\ }%
  \bibfield{journal}{%
  \bibinfo {journal} {Physical Review Letters}\ }%
  \textbf{\bibinfo {volume} {76}},\ \bibinfo {pages} {404} (\bibinfo {year}
  {1996})%
  \bibAnnoteFile{NoStop}{josephson-strogatz}%
\bibitem{daniels}%
  \BibitemOpen
  \bibfield{author}{%
  \bibinfo {author} {\bibfnamefont{B.~C.}\ \bibnamefont{Daniels}},\ }%
  \bibfield{journal}{%
  \bibinfo {journal} {Published Online}}%
   (\bibinfo {year} {2005}),\
  \url{http://go.owu.edu/~physics/StudentResearch/2005/BryanDaniels/kuramoto_p%
aper.pdf}%
  \bibAnnoteFile{NoStop}{daniels}%
\bibitem{strogatz-delay}%
  \BibitemOpen
  \bibfield{author}{%
  \bibinfo {author} {\bibfnamefont{M.~K.~Stephen}\ \bibnamefont{{Yeung}}}\ and\
  \bibinfo {author} {\bibfnamefont{S.~H.}\ \bibnamefont{{Strogatz}}},\ }%
  \bibfield{journal}{%
  \bibinfo {journal} {Physical Review Letters}\ }%
  \textbf{\bibinfo {volume} {82}},\ \bibinfo {pages} {648} (\bibinfo {year}
  {1999})%
  \bibAnnoteFile{NoStop}{strogatz-delay}%
\end{thebibliography}

%

\end{document}